\documentclass[aps,twocolumn,preprintnumbers,amsmath,amssymb]{revtex4}
\usepackage{amsfonts}
\usepackage{amsmath}
\usepackage{amssymb}
\usepackage{graphicx}
\usepackage{textcomp}
\begin{document}
\title{Frequency Comb Assisted Diode Laser Spectroscopy \\ for Measurement of Microcavity Dispersion}
\author{P.~Del'Haye$^1$, O.~Arcizet$^1$, M.~L.~Gorodetsky$^{1,2}$, R.~Holzwarth$^1$, T.~J.~Kippenberg$^{1,3}$} 
\email{tobias.kippenberg@epfl.ch}
\affiliation{$^1$Max-Planck-Institut f\"{u}r Quantenoptik, 85748 Garching, Germany}
\affiliation{$^2$Department of Physics, Moscow State University, Moscow, 119899, Russia}
\affiliation{$^3$\'{E}cole Polytechnique F\'{e}d\'{e}rale de Lausanne (EPFL), CH 1015, Lausanne, Switzerland}

\begin{abstract}
While being invented for precision measurement of single atomic transitions, frequency combs have also become a versatile tool for broadband spectroscopy in the last years. In this paper we present a novel and simple approach for broadband spectroscopy, combining the accuracy of an optical fiber-laser-based frequency comb with the ease-of-use of a tunable external cavity diode laser. This scheme enables broadband and fast spectroscopy of microresonator modes and allows for precise measurements of their dispersion, which is an important precondition for broadband optical frequency comb generation that has recently been demonstrated in these devices. Moreover, we find excellent agreement of measured microresonator dispersion with predicted values from finite element simulations and we show that tailoring microresonator dispersion can be achieved by adjusting their geometrical properties.
\end{abstract}
\maketitle
\revised{}

Spectroscopy has attracted attention of generations of scientists, starting with Fraunhofer's discovery of dark lines in the sun spectrum in 1814 and the work of Kirchhoff and Bunsen in 1859. Within the last decade, the invention of the optical frequency comb has revolutionized the field of spectroscopy and allowed measurements with previously unattainable precision \cite{Udem2002,Cundiff2003,YeCundiff2005}.
However, despite significant advances \cite{Diddams2007,Keilmann2004,Schliesser2005,Coddington2008,Mandon2009}, the use of frequency combs for precise broadband spectroscopy has remained challenging owing to low optical power per comb line and the difficulty to resolve features that are smaller than the combs free spectral range. 

Here, we present a novel and easy-to-use scheme for fast, broadband and precise spectroscopy combining the accuracy of an optical frequency comb with the broad bandwidth, high power and tunability of a mode-hop-free external cavity diode laser. We achieve sub-MHz-resolution over a bandwidth exceeding 4 THz in the 1550-nm range at scanning speeds up to 1 THz per second. As application of the scheme we present measurements of the transmission spectra of ultra-high-Q microcavities, which exhibit spectrally narrow absorption features ($<10$ MHz).  Our novel technique allows us to determine microcavity dispersion over a broad wavelength region for the first time. The combination of high resolution, extremely high scanning speed and ease of use makes this technique promising for a wide range of applications in photonics, sensing, photonic device or laser characterization.

Using optical frequency combs for broadband spectroscopy has so far been carried out in several ways. A powerful approach  - that uses all comb modes simultaneously - is based on the use of two frequency combs, with slightly different repetition rates. This multi-heterodyne technique allows to map large optical bandwidth into the radio frequency signals \cite{Keilmann2004,Schliesser2005,Coddington2008}. While highly accurate Hz level spectroscopy using simultaneously all comb modes has been achieved, the major drawback of this method is that it cannot resolve features below the repetition rate of the comb. This requires to scan the offset frequency (or the repetition rate), losing the advantage of the parallelism. In addition, the weak power per comb mode makes this approach unsuitable in many instances and moreover it requires two optical frequency combs, making it inaccessible for most laboratories and university settings. 

\begin{figure}[ptbh]
\begin{center}
\includegraphics[width=1\linewidth]{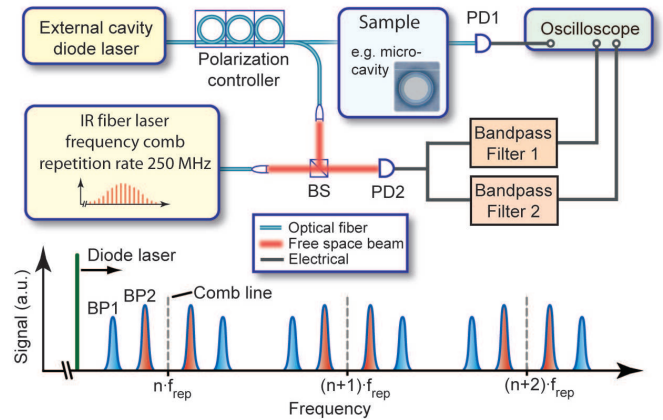}
\end{center}
\caption{\textbf{Measurement scheme.} A tunable diode laser is used to generate a ``running'' beat note with a fiber laser frequency comb. The beat note is detected by a photodiode (PD2) and sent through two band-pass filters (BP) to an oscilloscope, generating calibration markers for precise determination of the diode laser frequency. Another part of the diode laser signal is simultaneously used for spectroscopy. BS denotes beam splitter.}
\end{figure}

Another approach has been the idea of imprinting the accuracy of an optical frequency comb to a transfer laser with higher intensity, which is already utilized since the first spectroscopy experiments with frequency combs and enabling determination of transition frequencies of single atomic and molecular resonances with unprecedented accuracy \cite{Cundiff2001,Udem2002}. While in this scheme, the transfer laser has to remain phase locked to a frequency comb or high finesse cavity, the latter makes it challenging to perform measurements over a large bandwidth within short timescales \cite{Schibli2005,Jost2002}. In contrast to this earlier work, the presented scheme employs a free-running diode laser that is calibrated ``on the fly'' at fixed frequency intervals by a frequency comb. Moreover, the fast data acquisition allows for precise measurement of temporally unstable spectral features, eg. the mode spectra of microcavities that are drifting with temperature.

\begin{figure}[ptbh]
\begin{center}
\includegraphics[width=1\linewidth]{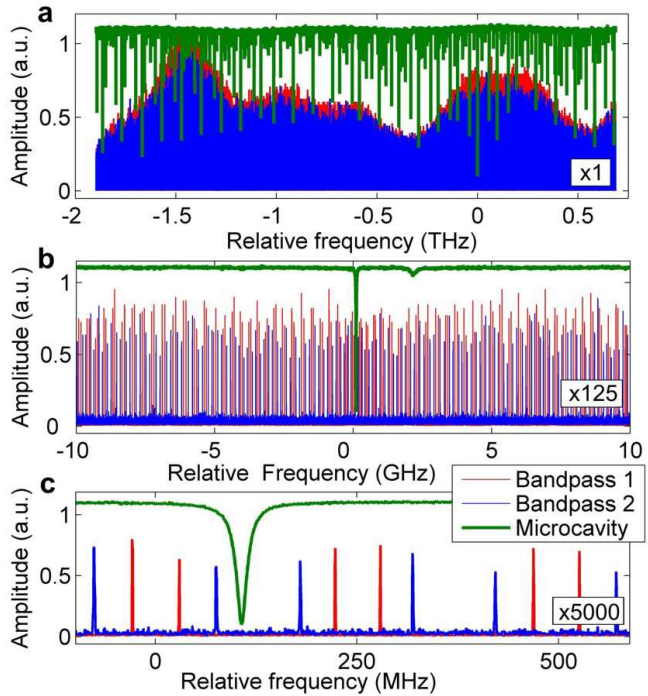}
\end{center}
\caption{\textbf{Frequency comb calibrated transmission spectrum.} The figure shows the transmission spectrum of a 400-$\mu$m-diameter silica microresonator with a free spectral range of 83 GHz and a typical linewidth of ca. 15 MHz. The data (green) and calibration peaks (blue, red) are recorded simultaneously for different frequency spans of 2.5 THz (a), 20 GHz (b) and 500 MHz (c). The presented configuration with two rf-filters at 30 MHz and 75 MHz leads to the generation of four calibration markers in each interval of one mode spacing (250 MHz) of the fiber laser comb.}
\end{figure}

\section{Results}

The experimental setup for the broadband precision spectroscopy is depicted in Fig 1. An external cavity diode laser with a tuning range of ca. 80 nm around 1550 nm (New Focus Velocity, short-term linewidth $< 300$ kHz) is used to generate a radio frequency (rf) beat note with an erbium-fiber-laser-based frequency comb (Menlo Systems GmbH), whose repetition rate $f_{\mathrm{rep}}=250$ MHz and carrier envelope offset frequency $f_{\mathrm{ceo}}$ are fully stabilized to a GPS disciplined hydrogen maser. Consequently, the modes of the frequency comb are fixed at $f_\mathrm{c} = f_{\mathrm{ceo}} + n \cdot f_{\mathrm{rep}}$ $(n \in \mathbb{N})$. The frequency comb modes are used to generate beat notes with the tunable diode laser, which are detected by a fast photodiode (PD2, 200 MHz bandwidth). When scanning the diode laser, ``running'' beat notes are generated, while particularly the beat notes with the two nearest comb lines at frequencies $f_{\mathrm{beat}}$ and $f_\mathrm{rep}-f_{\mathrm{beat}}$ are of interest. By sending the beat note signal to a narrowband band-pass filter (full width half maximum $\Gamma \approx 1$ MHz) and recording the output with an oscilloscope, it is possible to obtain calibration markers every time a beat note falls within the bandpass filter frequency centered around $f_\mathrm{BP}$, with $0 < f_\mathrm{BP}< f_\mathrm{rep} / 2$. More precisely, a frequency reference marker is recorded when the diode laser frequency equals to $f_\mathrm{D} = f_{\mathrm{ceo}} + n \cdot f_{\mathrm{rep}} \pm f_{\mathrm{BP}}$. Using the ``peak-detect'' sampling mode of the employed oscilloscope enables recording of the calibration marker without actually resolving the beat note frequency. Additionally, to increase the number of reference markers, the electrical beat note signal is split up and sent to two different bandpass filters at frequencies of 30 MHz and 75 MHz (passband width $\Gamma \approx 1$ MHz). This leads to the generation of four calibration peaks within each interval of one repetition rate ($f_{\mathrm{rep}}=250$ MHz), located at $\pm 30$ MHz and $\pm 75$ MHz with respect to each comb line (cf. Fig 2). Note that the exact center frequency of the bandpass filters can be directly inferred from the measurement and does not have to be known a priori. The calibration peaks created by the two rf-filters are recorded using two ports of a four-channel oscilloscope supporting 10 million data points per channel, leading to a fundamental resolution of 370 kHz per datapoint for a 30 nm scan range. At the same time, the remaining channels of the oscilloscope can be used to record a spectrum of interest using a photodiode and the scanning diode laser. One measurement gives rise to more than 64,000 calibration peaks with a sufficient signal-to-noise ratio to be detected with an automated data analysis routine. The instantaneous frequency of the diode laser can be interpolated using these calibration markers. Experiments have shown that a spline-interpolation (polynomial) of the diode laser frequency scan leads to slightly better results than a simple linear model, leading to a resolution of $\approx 1$ MHz. This resolution is determined by the scanning behavior of the diode laser and has been determined by measuring the well defined frequency of a second diode laser that has been phase locked to the frequency comb. On the other hand, the maximum useful diode laser scanning speed $v_\mathrm{max}$ is determined by the transmission behavior of the employed rf-filters. It can be estimated by the constraint that the time needed by the beat note to cross the band-pass filter has to be smaller than the filter's response time ($\Gamma / v_\mathrm{max} = 1 / \Gamma$). Using a band-pass filter with a bandwidth of $\Gamma \approx 1\ \mathrm{MHz}$ yields a maximum scanning speed of $v_\mathrm{max} = 1 / \Gamma^2 \approx 1\ \mathrm{THz}/\mathrm{s}$ ($8\ \mathrm{nm} / \mathrm{s}$ at 1550 nm). This agrees well with experimental observations, showing that the calibration markers cannot be recorded with a sufficient signal-to-noise ratio at faster scanning speeds. Additionally, at higher scanning speeds one has to take into account systematic errors as a consequency of the transient oscillation of the bandpass filters. The same limitations apply to the part of the diode laser used for the actual measurement. In case of measuring spectral features that are smaller than the filter bandwidth, $\Gamma$ has to replaced by the linewidth of these spectral features. It has to be noted that the presented measurement scheme requires an additional measurement of at least one point in the spectrum with an accuracy higher than the repetition rate of the employed frequency comb for calibrating the absolute position of the recorded spectrum. Alternatively, the ambiguity of the recorded spectrum can also be resolved by recording another dataset with a slightly different repetition rate of the frequency comb.

\begin{figure}[ptbh]
\begin{center}
\includegraphics[width=1\linewidth]{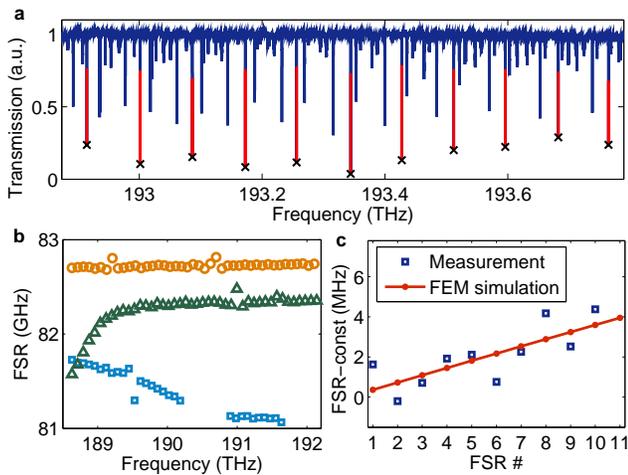}
\end{center}
\caption{\textbf{Microcavity mode spectrum and dispersion. a,} High-Q mode family of a toroidal silica microresonator with a linewidth of ca. 10 MHz and a mode spacing of ca. $83$ GHz. \textbf{b}, Evolution of the free spectral range of different mode families in an 800-$\mu$m-diameter resonator. Orange circles depict the fundamental mode family. The outliers and the strongly inclined part of the green dataset (triangles) is a result of mode crossings with modes from other mode families. \textbf{c}, Evolution of subsequent free spectral ranges of the mode family in panel a, starting with the FSR located at the lowest frequency. The solid line shows data from finite element simulations of a resonator with the same geometry. The measured value of $(370 \pm 110) \ \frac{\mathrm{kHz}}{\mathrm{FSR}}$ is in good agreement with the simulated value of $350 \ \frac{\mathrm{kHz}}{\mathrm{FSR}}$.}
\end{figure}

As a proof-of-concept experiment, the scanning diode laser spectroscopy scheme is used to measure mode spectra of ultra-high-Q optical microcavities \cite{Armani2003}. This precision measurement allows to determine the dispersion of optical microcavity modes by exploiting all advantages of the spectroscopy scheme discussed before. Recently, it has been shown that ultra-high-Q microcavities allow to generate frequency combs with a bandwidth of more than 500 nm via four-wave mixing \cite{Del'Haye2007,Del'Haye2008,Savchenkov2008a}. The total bandwidth of these combs is limited by the dispersion of the microcavity, which render the resonator modes non-equidistant. Thus, understanding and measurement of microcavity dispersion is an essential step towards octave spanning comb generation in microcavities, required to implement self referencing \cite{Udem2002,Telle1999}. The dispersion of microcavities leads to a walk-off between subsequent free spectral ranges (FSRs) of a microcavity and is expected to be smaller than $1 \ \mathrm{MHz}/\mathrm{FSR}$ (cf. measurement of dispersion in a CaF$_2$ resonators using a whitelight source and a tunable optical filter \cite{Savchenkov2008}). Measurement of microcavity dispersion has been out of reach of existing spectroscopy methods \cite{Schibli2005,Jost2002} owing to the large free spectral range of up to 1 THz, the small dispersion values ($< 1 \ \mathrm{MHz}/\mathrm{FSR}$), narrow optical linewidths ($< 10$ MHz) and thermal drifts of microcavity modes ($\approx 10 \ \mathrm{GHz} / \mathrm{K}$), requiring fast acquisition of the mode spectrum. While precise dispersion measurements of macroscopic Fabry-Perot cavities have already been carried out in earlier experiments by stepwise measuring the distance between resonator modes and using a Vernier effect \cite{Thorpe2005,Schliesser2006a}, especially the thermal drift in microresonators requires faster acquisition of the position of the cavity modes.

It is important to distinguish the present approach to dispersion measurements from previous approaches based on frequency combs and applied to characterize mirrors of Fabry-Perot resonators \cite{Thorpe2005,Schliesser2006a}. In the latter work the repetition rate of the frequency comb and those of the Fabry-Perot were almost identical, allowing carrying out dispersion measurements whereby the (closely spaced) individual comb modes did not have to be resolved. In the present setting, where the individual resonances are many integers of the repetition rate apart, and where dispersion manifests itself only in the transmitted power of the cavity, it would be necessary to resolve the closely spaced individual comb modes making the previously demonstrated technique unsuitable for measurement of the present microcavity dispersion.

Figure 3a shows a calibrated mode spectrum of a microtoroid with 800 $\mu$m diameter and a corresponding 83-GHz mode spacing. Several different mode families can be seen, while the one with the highest optical quality factor of $Q \approx 2 \times 10^7$ is marked with crosses. The modes of this spectrum have been fitted with a Lorentzian curve and used to calculate the evolution of the FSR of a microcavity. Figure 3b shows the measured dispersion data of different mode families of the same microcavity. The difference in FSR between different mode families is on the order of 1 GHz, resulting from different spatial mode structure within the toroid. Note that some mode positions are strongly affected by mode crossings with other families \cite{Carmon2008}, leading to a locally increased dispersion. The evolution of the FSR of the mode with the highest optical Q-factor in this cavity is depicted in Fig. 3c and is in good agreement with finite element simulations that have been carried out using the fundamental modes of a microtoroid with the same geometry. In particular our results reveal that the dispersion is anomalous (which would allow for formation of optical solitons \cite{MOLLENAUER1980}).

\begin{figure}[hptb]
\begin{center}
\includegraphics[width=\linewidth]{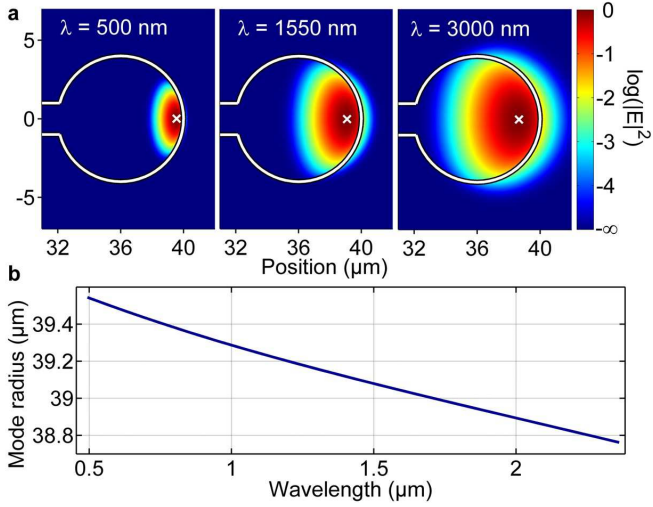}
\end{center}
\caption{\textbf{Geometric dispersion of a microresonator. a,} Optical TE modes in a 40-micron-radius microtoroid at different wavelengths. The colorcode represents the electrical field intensity $|E|^2$ in logarithmic scale. \textbf{b}, Wavelength dependence of the maximum of the optical mode (cf. crosses in Figure 4a) as a function of the whispering gallery mode wavelength.}
\end{figure}

For a comparison with the experimental results, numerical simulations of the mode spectrum of microtoroids have been carried out using a commercially available finite element simulation software as described in Ref. \cite{Oxborrow2007}. These simulations allow to calculate resonance frequencies of whispering gallery modes for a given axisymmetric geometry, mode number and refractive index. To calculate the dispersion of microresonators one has to distinguish between the contribution of the geometry dependent waveguide dispersion and the intrinsic material dispersion, which can be described by the Sellmeier equation for fused silica. To add the material dispersion contribution to our simulations, we have chosen an iterative approach, in which the refractive index is stepwise adjusted to new values infered from the resonance frequency of the preceding iteration. The impact of the geometric dispersion on the mode profile can be seen in Fig. 4. The ``center of mass'' of the optical mode is moving to the resonator rim for decreasing wavelength, corresponding to an increasing optical pathlength with frequency (which is referred to as normal dispersion). On the other hand, the anomalous material dispersion of fused silica for wavelengths above 1300 nm is counteracting the geometric dispersion of the resonator. To derive the group delay dispersion GDD of a microresonator (which determines the variation of the FSR), we define a frequency dependent optical pathlength for one roundtrip in a circular resonator by: 
\begin{equation}
	L(\omega)= 2 \pi R_0 \cdot n_0 + c \cdot \frac{\mathrm{d} \varphi}{\mathrm{d} \omega} \ ,
\end{equation}
with $R_0$ and $n_0$ being the mode radius and refractive index of the resonator at frequency $\omega_0$, $c$ being the speed of light in vacuum and $\frac{\mathrm{d} \varphi}{\mathrm{d} \omega}$ being the additional group delay induced by material and geometric dispersion. Thus, the group delay dispersion GDD is given as
\begin{equation}
	\mathrm{GDD}(\omega) = \frac{\mathrm{d}^2 \varphi}{\mathrm{d} \omega^2} = \frac{\mathrm{d}}{\mathrm{d} \omega} \frac{L(\omega)}{c} = \frac{1}{2 \pi} \cdot \frac{\mathrm{d}}{\mathrm{d} \nu} \frac{1}{FSR} \ .
\end{equation}

The total GDD of microresonators with different sizes, including geometric- and material-dispersion contributions, is depicted in Fig. 5a. Different curves correspond to different radii of the toroidal resonator. The minor radius (determining the toroid's cross-section) is 4 micron for all simulations. As a result of the geometric dispersion, the zero-dispersion wavelength $\Lambda_0$ is shifted to higher values for smaller resonators (cf. Fig. 5b), enabling precise control of the microcavities dispersion in the vicinity of the pump laser for frequency comb generation (cf. Ref. \cite{Agha2007}). Another approach to calculate the dispersion of a toroidal microresonator is to use an analytical approximation. As far as whispering gallery modes adjacent to the equatorial plane are concerned, toroids with large radius $R$ and small radius $r$ are very close to a spheroid with semiaxis $a=R$ and $b=\sqrt{Rr}$ with eigenfrequencies \cite{Gorodetsky2006, Gorodetsky2007} determined by:  
\begin{align}
\tilde y &\equiv \frac{n\omega_0 a}{c} \simeq 
 \ell-\alpha_q\left(\frac{\ell}{2}\right)^{\frac{1}{3}}+\frac{2p(a-b)+a}{2b} \nonumber\\
 &-\frac{P n}{\sqrt{n^2-1}} \label{freq2} +\frac{3\alpha_q^2}{20}\left(\frac{\ell}{2}\right)^{-\frac{1}{3}}\nonumber\\ &-\frac{\alpha_q}{12}\left(\frac{2p(a^3-b^3)+a^3}{b^3}+\frac{2n^3P(2P^2-3)}{(n^2-1)^{\frac{3}{2}}}\right)\left(\frac{\ell}{2}\right)^{-\frac{2}{3}}\nonumber,
\end{align}
where $\ell$ is the mode number, $p$ is the number of variations in meridional direction, $P=1$ or $P=1/n_0^2$, for the TE or TM modes correspondingly, $\alpha_q$ is the root of the order $q$ of the Airy function ($p=0$ and $q=1$ for fundamental modes).  Using this equation all dispersion parameters of a toroid may be approximated, and in particular: 
\begin{align}
&GDD(\omega)=\frac{\partial}{\partial\omega}\left(\frac{\partial\omega}{\partial \ell}\right)^{-1}\\
&=\frac{n^2a^2}{c^2}\Big(\frac{\partial \tilde y}{\partial \ell}\Big)^{-3}\left[\frac{1}{\tilde y}\Big(\frac{\partial \tilde y}{\partial \ell}\Big)^2\frac{\lambda^2}{n} \frac{\partial^2 n}{\partial\lambda^2}-\frac{\partial^2\tilde y}{\partial\ell^2}\Big(1-\frac{\lambda}{n}\frac{\partial n}{\partial\lambda}\Big)^2\right]\nonumber.
\end{align}
The zero dispersion wavelength $\Lambda_0$ is thus determined by the following expression which allows to obtain $\ell$ and hence from (\ref{freq2}) optimal dimensions for a given $\Lambda_0$:
\begin{align}
{\tilde y}\frac{\partial^2\tilde y}{\partial\ell^2}\Big/\Big(\frac{\partial \tilde y}{\partial \ell}\Big)^2=\frac{\lambda^2}{n} \frac{\partial^2 n}{\partial\lambda^2}\Big/\Big(1-\frac{\lambda}{n}\frac{\partial n}{\partial\lambda}\Big)^2.
\end{align}
Taking only the leading terms in (\ref{freq2}) we can deduce a rather crude analytical approximation for $\ell$ which ignores additional dispersion in very flattened spheroids with large $a$:
\begin{align}
\ell&=2\left[\frac{\alpha_q\Big(1-\frac{\lambda}{n}\frac{\partial n}{\partial\lambda}\Big)^2}{9\frac{\lambda^2}{n}\frac{\partial^2n}{\partial\lambda^2}}\right]^{3/2}.
\end{align}

\begin{figure}[hptb]
\begin{center}
\includegraphics[width=\linewidth]{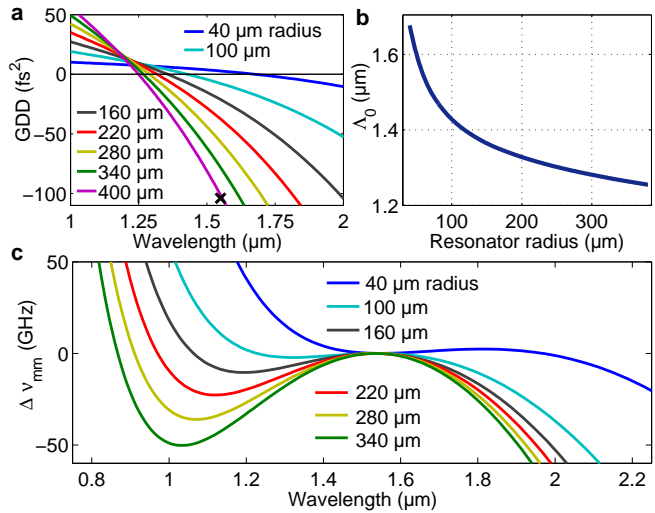}
\end{center}
\caption{\textbf{Dispersion of a microresonator. a,} Group delay dispersion for resonators of different size. The black cross marks the measured GDD of -104 fs$^2$ in a 400-micron-radius toroid. \textbf{b}, Dependence of the zero-dispersion wavelength $\Lambda_0$ of a microresonator on the radius. \textbf{c}, Mismatch between comb and microcavity modes for a pump laser at 1550 nm. The mode spacing of the comb modes is defined by the cold cavities FSR around the pump wavelength.}
\end{figure}

A more intuitive way to specify the dispersion of a microresonator pertaining to optical frequency comb generation via four-wave mixing \cite{Del'Haye2007} is the mismatch between the generated comb modes $\nu_\mathrm{comb}$ and the respective microcavity modes $\nu_\ell$ (with $\ell$ being the angular mode number). Assuming that the central comb mode correponds to a certain microresonator mode at frequency $\nu_{\ell_0}$, the frequency mismatch is given by
\begin{equation}
	\Delta \nu_\mathrm{mm} (\ell) = \nu_\mathrm{comb} - \nu_\ell = \nu_{\ell_0} + \Delta \ell \cdot \nu_\mathrm{rep} - \nu_\ell \ ,
\end{equation}
with $\nu_\mathrm{rep}$ being the mode spacing of the generated frequency comb, which is determined by the microresonator mode structure in the vicinity of the central mode at $\nu_{\ell_0}$ and $\Delta \ell = \ell - \ell_0$ is an integer number defining the mode number offset to the mode of interest. Figure 5c shows the wavelength dependent frequency mismatch $\Delta \nu_\mathrm{mm}$ for resonators of different sizes at a pump wavelength of 1550 nm. Even in the case of a 40-micrometer-diameter resonator with a GDD of ca. 10 fs$^2$, the mismatch between comb modes and microcavity modes can reach more than 10 GHz at wavelength below 1.3 $\mu$m, corresponding to a mismatch of several hundred linewidths. This is a surprising result, since optical frequency combs in such a resonator with the same pump wavelength of 1550 nm have been observed down to less than 1.1 $\mu$m. An explanation for this effect can be given by a self compensation of the microcavity dispersion via the optical Kerr-effect \cite{Kippenberg2004a}. Indeed, the Kerr-nonlinearity induced frequency shift $\Delta \nu$ in an ultra-high-Q microresonator can exceed several tenth of gigahertz as a result of the huge power enhancement of the cavity (with resonator finesses exceeding $10^5$). This frequency shift depends on the intracavity power at the spatial position of a certain optical mode and can be estimated as $\Delta \nu = \frac{\Delta n}{n_0} \cdot \nu_0$ with $\Delta n = n_2 \cdot I_\mathrm{res}$ being the intensity dependend refractive index change induced by the Kerr-nonlinearity. Calculating the intensity in the resonator $I_\mathrm{res}$ at a launched power of $P_\mathrm{in} = 200$ mW, we can derive an approximate expression based on a fourier limited pulse in the cavity (i.e. soliton case):

\begin{equation}
\Delta \nu = \frac{\nu_0}{n_0^2} \cdot n_2 \cdot \frac{c}{4 \pi^2 \cdot R \cdot A_\mathrm{eff}} \frac{N}{\kappa} P_\mathrm{in} \ .
\end{equation}

For an effective mode area of $A_\mathrm{eff} \approx (1 \ \mathrm{\mu m})^2$, a resonator radius of $R \approx 100 \ \mathrm{\mu m}$, a linewidth of the resonator modes of $\kappa \approx 10$ MHz, and for $N \approx 20$ comb lines contributing to the resonance shift this equates to a nonlinear frequency shift of $\Delta \nu \approx 60 \ \mathrm{GHz}$ (using $n_0 =1.44$ and $n_2 = 2.2 \cdot 10^{-20} \frac{\mathrm{m^2}}{\mathrm{W}}$ as the linear and nonlinear refractive index of fused silica).

\section{Discussion}

In conclusion we have presented a novel scheme for broadband precision spectroscopy that enables the measurement of spectral features with sub-MHz-linewidth at previously unattainable measurement speed exceeding 1 THz per second and over a bandwidth of more than 4 THz. This scheme is used to determine high-resolution spectra of microresonator modes, allowing the measurement of their modal dispersion for the first time. Additionally, finite element simulations of microcavity dispersion have been carried out, which are in good agreement with the experimental results. The presented spectroscopy scheme enables a wide range of applications including characterization of photonic devices, gas spectroscopy and molecular sensing. Indeed, we believe that the simplicity of the scheme that relies on widely employed tunable diode lasers, will allow to significantly extend the number of laboratories in which optical frequency comb based spectroscopy can be carried out and make this technique widely available. Moreover the need to calibrate photonic devices may be served well with our technique.

\bibliographystyle{nature}

\section{Acknowledgements}

We thank Mark Oxborrow for providing templates for FEM simulation of whispering gallery modes in toroidal microresonators. T.J.K. acknowledges support via an Independent Max Planck Junior Research Group. This work was funded as part of a Marie Curie Excellence Grant (RG-UHQ) and the DFG funded Nanosystems Initiative Munich (NIM). We thank the Max Planck Institute of Quantum Optics and P. Gruss for continued support. M.L.G acknowledges support from the Alexander von Humboldt Foundation.

\end{document}